\documentclass{webofc}
\usepackage[varg]{txfonts}   
\begin{document}
\title{\boldmath
Inclusive Hadroproduction of $P$-wave Heavy Quarkonia in pNRQCD}

\author{\firstname{Hee Sok} \lastname{Chung}\inst{1,2}\fnsep\thanks{\email{heesok.chung@tum.de}} 
}

\institute{Physik-Department, Technische Universit\"at M\"unchen,
James-Franck-Str. 1, 85748 Garching, Germany
\and
Excellence Cluster ORIGINS,
Boltzmannstrasse 2, D-85748 Garching, Germany}

\abstract{
We compute NRQCD long-distance matrix elements that appear in the inclusive
production cross sections of $P$-wave heavy quarkonia in the framework of
potential NRQCD. 
The formalism developed in this work applies to
strongly coupled charmonia and bottomonia. 
This makes possible the determination of color-octet NRQCD
long-distance matrix elements without relying on measured cross section data, 
which has not been possible so far. 
We obtain results for inclusive production 
cross sections of $\chi_{cJ}$ and $\chi_{bJ}$ at the LHC, which are in good
agreement with measurements. 
}

\maketitle

\section{Introduction}
\label{intro}
Hard processes involving heavy quarkonia are considered good probes of
perturbative and nonperturbative aspects of QCD~\cite{Brambilla:2004wf, 
Brambilla:2010cs, Bodwin:2013nua, Brambilla:2014jmp}. 
Especially, inclusive production cross sections of heavy quarkonia are 
studied extensively in hadron colliders. 
A good theoretical understanding of the quarkonium production mechanism is
therefore important in deciphering QCD interactions from collider experiments. 

The effective field theory nonrelativistic QCD 
(NRQCD)~\cite{Caswell:1985ui, Bodwin:1994jh} 
provides a factorization formalism for hard processes involving heavy
quarkonia, which include inclusive production cross sections at large
transverse momentum. 
NRQCD utilizes the separation
of the scale of the heavy quark mass $m$ from the scales $mv$ and $mv^2$, where
$v \ll 1$ is the relative velocity of the heavy quark $Q$ and the antiquark
$\bar Q$ inside the heavy quarkonium. The NRQCD factorization
formula for inclusive production rate of a quarkonium 
${\cal Q}$ can be written as 
\begin{equation}
\label{eq:fac}
\sigma_{{\cal Q}+X} = \sum_n \hat{\sigma}_{Q\bar{Q} (n)+X} 
\langle {\cal O}^{\cal Q} (n) \rangle,
\end{equation}
where the $\hat{\sigma}_{Q\bar{Q} (n)+X}$ are
short-distance cross sections (SDCSs) for inclusive production of a 
$Q \bar Q$ pair 
in a specific color and angular momentum state $n$, while 
the long-distance matrix elements (LDMEs) 
$\langle {\cal O}^{\cal Q} (n)\rangle$ 
correspond to the nonperturbative probabilities for the $Q \bar Q$ to 
evolve into the quarkonium $\cal Q$. 
The LDMEs have known scalings in $v$, so that the sum (\ref{eq:fac}) can
be organized in powers of $v$, and is typically truncated at a desired 
accuracy in $v$. 
While the short-distance
cross sections can be computed in perturbative QCD as series in the
strong coupling $\alpha_s$, the LDMEs must be determined
nonperturbatively. 

While it has been known how to compute the color-singlet LDMEs, 
determinations of the color-octet LDMEs have usually been done by 
relying on measured cross section data, because it has not been known so far 
how to compute them from first principles. This approach led to
inconsistent sets of LDME determinations that do not provide a comprehensive
description of the important observables involving inclusive production of
heavy quarkonia~\cite{Chung:2018lyq}. 
Therefore, it is much desirable to be able to compute 
color-octet LDMEs from first principles. 

In the strongly coupled regime 
(characterized by the hierarchy $\Lambda_{\rm QCD} \gg mv^2$), 
the potential NRQCD (pNRQCD) effective field theory~\cite{
Brambilla:1999xf, Pineda:1997bj, Brambilla:2004jw}
which is obtained by further integrating out the scale $mv$, 
provides expressions for the LDMEs 
for decays of heavy quarkonia into light particles 
in terms of quarkonium wavefunctions and gluonic 
correlators~\cite{Brambilla:2001xy, Brambilla:2002nu, 
Brambilla:2020xod}. 
This greatly enhances the predictive power of the NRQCD
factorization formalism, as the wavefunctions can be computed by solving
the nonrelativistic Schr\"odinger equation (see, for example,
refs.~\cite{Chung:2020zqc, Chung:2021efj}), 
and the gluonic correlators are
universal quantities that do not depend on the heavy quark flavor or the radial
excitation, and in principle can be computed in lattice QCD. 
It has been anticipated that
a similar calculation could be done for the LDMEs for inclusive production
cross sections. This formidable task has been accomplished for the first time
in ref.~\cite{Brambilla:2020ojz} for the production of strongly coupled 
$P$-wave heavy quarkonia, 
and a general formalism for strongly coupled quarkonia of arbitrary 
quantum numbers has been established in 
ref.~\cite{Brambilla:2021abf}. Similarly to the case
of decay LDMEs, the results in refs.~\cite{Brambilla:2020ojz,
Brambilla:2021abf} provide expressions for the production
LDMEs in terms of quarkonium wavefunctions 
and universal gluonic correlators. 
These results greatly reduce
the number of nonperturbative unknowns and enhances the predictive power of the
NRQCD factorization formalism for inclusive production of heavy quarkonia. 

\section{\boldmath Inclusive production of $\chi_{cJ}$ and $\chi_{bJ}$ in
pNRQCD}
\label{sec-2}

The NRQCD factorization formula for the inclusive production rate of 
$\chi_{QJ}$ ($Q=c$ or $b$) at leading order in $v$ is given
by~\cite{Bodwin:1994jh}
\begin{equation}
\label{eq:pwave_fac}
\sigma_{\chi_{QJ}} = 
(2 J+1) \left[ \hat{\sigma}_{Q \bar{Q}(^3P_J^{[1]})} 
\langle {\cal O}^{\chi_{Q0}} (^3P_0^{[1]}) \rangle
+ 
\hat{\sigma}_{Q \bar{Q}(^3S_1^{[8]})} 
\langle {\cal O}^{\chi_{Q0}} (^3S_1^{[8]}) \rangle \right], 
\end{equation}
where the LDMEs are defined by~\cite{Bodwin:1994jh,
Nayak:2005rt, Nayak:2005rw, Nayak:2006fm}
\begin{subequations}
\begin{align}
\langle {\cal O}^{\chi_{Q0}} (^3P_0^{[1]}) \rangle
&= 
\langle \Omega | 
\frac{1}{3} \chi^\dag \left(-\frac{i}{2} \overleftrightarrow{{\bf D}} \cdot
{\boldsymbol \sigma} \right) \psi
\, {\cal P}_{\chi_{Q0}({\bf P}={\bf 0})} \, \psi^\dag \left(-\frac{i}{2}
\overleftrightarrow{{\bf D}} \cdot {\boldsymbol \sigma} \right) \chi | \Omega \rangle,
\\
\langle {\cal O}^{\chi_{Q0}}({}^3S_1^{[8]})\rangle &= 
\langle \Omega | \chi^\dag \sigma^i T^a \psi \Phi_\ell^{\dag ab} \, {\cal
P}_{{\chi_{Q0}}({\bf P}={\bf 0})} \, \Phi_\ell^{bc} \psi^\dag
\sigma^i T^c \chi | \Omega \rangle.
\end{align}
\end{subequations}
Here, $|\Omega \rangle$ is the QCD vacuum, 
$\psi$ and $\chi$ are Pauli spinor fields that annihilate and create a 
heavy quark and
antiquark, respectively, ${\bf D} = {\boldsymbol \nabla}-i g {\bf A}$ is the
gauge-covariant derivative, $\chi^\dag \overleftrightarrow{\bf D} \psi
= \chi^\dag {\bf D} \psi - ({\bf D} \chi)^\dag \psi$, 
${\bf A}$ is the gluon field, $\boldsymbol \sigma$ is a Pauli matrix, 
$T^a$ is a color matrix, and the operator ${\cal P}_{\chi_{Q0}({\bf P})}$ projects onto 
states that include the quarkonium $\chi_{Q0}$ with momentum ${\bf  P}$. 
The gauge-completion Wilson line defined by 
$\Phi_\ell 
= {\cal P} \exp \left[ -i g \int_0^\infty d \lambda \ell \cdot 
A^{\rm adj}(\ell \lambda) \right]$, 
where ${\cal P}$ is path ordering, $A^{\rm adj}$ is the gluon field in the
adjoint representation, and $\ell$ is an arbitrary direction, 
is inserted to ensure the gauge invariance of the color-octet 
LDME~\cite{Nayak:2005rt, Nayak:2005rw, Nayak:2006fm}. 
We have used the heavy-quark spin symmetry to write eq.~(\ref{eq:pwave_fac})
in terms of the $\chi_{Q0}$ LDMEs~\cite{Bodwin:1994jh}. 
An analogous formula holds for the production of $h_Q$. 

The pNRQCD expression for the production LDMEs can be written
as~\cite{Brambilla:2021abf} 
\begin{align}
\label{eq:ldmes_pnrqcd}
\langle \Omega | {\cal O}^{{\cal Q}} (n) | \Omega \rangle
&= \frac{1}{\langle {\bf P} ={\bf 0} | {\bf P}={\bf 0} \rangle} 
\int d^3x_1 d^3x_2 d^3x'_1 d^3x'_2 \, 
\phi_{\cal Q}^{(0)} ({\bf x}_1-{\bf x}_2) 
\nonumber\\
& \hspace{3ex} \times \left[ - V_{{\cal O}(n)} ({\bf x}_1,
{\bf x}_2;{\boldsymbol \nabla}_1,{\boldsymbol \nabla}_2)
\delta^{(3)} ({\bf x}_1-{\bf x}_1') \delta^{(3)} ({\bf x}_2-{\bf x}_2') 
\right] \phi_{\cal Q}^{(0)*} ({\bf x}_1'-{\bf x}_2'), 
\end{align}
where $V_{{\cal O}(n)}$ is a contact term, and 
$\phi_{\cal Q}^{(0)} ({\bf x}_1-{\bf x}_2)$ is the unit-normalized 
quarkonium wavefunction at leading order in $v$. 
The wavefunction is an eigenfunction of the pNRQCD Hamiltonian, 
which is given at leading order in $v$ by  
\begin{equation}
-\frac{\nabla_1^2}{2 m} -\frac{\nabla_2^2}{2 m} 
+ V^{(0)} ({\bf x}_1-{\bf x}_2), 
\end{equation}
where $V^{(0)}$ is the static potential. 
The expression in 
eq.~(\ref{eq:ldmes_pnrqcd}) is valid at leading order in $v$, 
up to corrections of relative order $1/N_c^2$, where $N_c$ is the number of
colors. 
We refer the readers to ref.~\cite{Brambilla:2021abf} for details of the
pNRQCD formalism for production LDMEs. 

The contact term $V_{{\cal O}(n)}$ can be computed in expansion in powers of
$1/m$. At leading nonvanishing orders in $1/m$, the contact terms for the 
LDMEs that appear in eq.~(\ref{eq:pwave_fac}) are given
by~\cite{Brambilla:2021abf}
\begin{subequations}
\begin{align}
- V_{{\cal O}(^3P_0^{[1]})} \big|_{P-{\rm wave}} &= 
- \frac{1}{3} \sigma^i \otimes \sigma^j 
N_c \nabla_{\bf r}^i \delta^{(3)} ({\bf r}) \nabla_{\bf r}^j, 
\\
- V_{{\cal O}(^3S_1^{[8]})} \big|_{P-{\rm wave}}&= 
- \sigma^k \otimes \sigma^k 
N_c \nabla_{\bf r}^i \delta^{(3)} ({\bf r}) \nabla_{\bf r}^j 
\frac{{\cal E}^{ij}}{N_c^2 m^2}, 
\end{align}
\end{subequations}
where ${\bf r} = {\bf x}_1-{\bf x}_2$, and 
we neglect contributions that vanish in (\ref{eq:ldmes_pnrqcd}) when 
$\phi_{\cal Q}^{(0)} ({\bf x}_1-{\bf x}_2)$ is in a $P$-wave state. 
The tensor ${\cal E}^{ij}$ is defined by 
\begin{equation}
\label{eq:Etensor}
{\cal E}^{ij} = 
\int_0^\infty dt\, t \; \int_0^\infty dt'\, t'\; \langle \Omega |
\Phi_\ell^{\dag ab} \Phi_0^{\dag ad} (0;t) g {E}^{d,i}(t) g {E}^{e,j}(t')
\Phi_0^{ec} (0;t') \Phi_\ell^{bc} | \Omega \rangle, 
\end{equation}
where $E^{i,a} (t) = 
E^{i,a} (t,{\bf 0}) = G^{i0,a} (t,{\bf 0})$ is the chromoelectric field, 
$G^{\mu \nu, a} T^a = G^{\mu \nu}$ is the gluon field strength tensor, 
and $\Phi_0(t;t') = {\cal P} \exp \left[ -i g \int_t^{t'} 
d \tau \, A_0^{\rm adj} (\tau,{\bf 0}) \right]$ 
is a Schwinger line. 

By using the results for the contact terms $V_{{\cal O}(n)}$ and
eq.~(\ref{eq:ldmes_pnrqcd}), we obtain the following expressions for the
production LDMEs in pNRQCD~\cite{Brambilla:2020ojz, Brambilla:2021abf}
\begin{subequations}
\begin{align}
\langle {\cal O}^{\chi_{Q0}} (^3P_0^{[1]}) \rangle
&=
\frac{3 N_c}{2 \pi} | R^{(0)}{}'(0)|^2, 
\\
\langle {\cal O}^{\chi_{Q0}}({}^3S_1^{[8]})\rangle &=
\frac{3 N_c}{2 \pi} | R^{(0)}{}'(0)|^2
\frac{\cal E}{9 N_c m^2}, 
\end{align}
\end{subequations}
where $R^{(0)}(r)$ is the radial wavefunction,
and ${\cal E} = \frac{3}{N_c} \delta^{ij} {\cal E}^{ij}$ is a 
gluonic correlator. 
We expect corrections to these expressions to be suppressed by $v^2$. Since
$v^2$ is comparable to $1/N_c^2$ for bottomonia, while $v^2$ is larger than
$1/N_c^2$ for charmonia, we take the uncertainties in the pNRQCD expressions
of the LDMEs to be of relative order $v^2$. 

The correlator ${\cal E}$ has a logarithmic scale dependence at one loop level,
which is given by 
\begin{equation}
\frac{d}{d \log \Lambda} {\cal E}(\Lambda) 
= 12 C_F \frac{\alpha_s}{\pi} + O(\alpha_s^2), 
\end{equation}
where $C_F = (N_c^2-1)/(2 N_c)$. 
This implies the following evolution equation for the LDMEs 
\begin{equation}
\frac{d}{d \log \Lambda} 
\langle {\cal O}^{\chi_{Q0}}({}^3S_1^{[8]})\rangle 
= \frac{4 C_F \alpha_s}{3 N_c \pi m^2} 
\langle {\cal O}^{\chi_{Q0}} (^3P_0^{[1]}) \rangle. 
\end{equation}
This agrees with the known one-loop evolution equation derived from
perturbative calculations in NRQCD~\cite{Bodwin:1994jh}. 
The consistency with NRQCD factorization at
two-loop level has been discussed in ref.~\cite{Brambilla:2021abf}.

Because the operator definition for ${\cal E}$ involves only gluon fields, 
it is independent of the radial excitation or the flavor of the heavy quark. 
Hence, the ratio of the color-singlet and 
color-octet LDMEs at leading order in $v$ given by 
\begin{equation}
\frac{m^2 \langle {\cal O}^{\chi_{Q0}}({}^3S_1^{[8]})\rangle}
{\langle {\cal O}^{\chi_{Q0}} (^3P_0^{[1]}) \rangle} 
= \frac{{\cal E}}{9 N_c}, 
\end{equation}
is universal for all $P$-wave quarkonium states. 
This, in turn, implies that a determination of ${\cal E}$ leads to
determination of both color-singlet and color-octet production LDMEs for 
all $P$-wave charmonium and bottomonium states. 

\subsection{\boldmath Inclusive hadroproduction of $\chi_{cJ}$}

We now present the phenomenological results for hadroproduction rates of 
$\chi_{cJ}(1P)$ ($J=1$, 2) at the LHC. 
We employ the SDCSs computed at next-to-leading order
accuracy in $\alpha_s$ from ref.~\cite{Wan:2014vka}, and take $\Lambda=1.5$~GeV
for the $\overline{\rm MS}$ scale associated with the color-octet LDME.

\begin{figure}[h]
\centering
\includegraphics[width=8cm,clip]{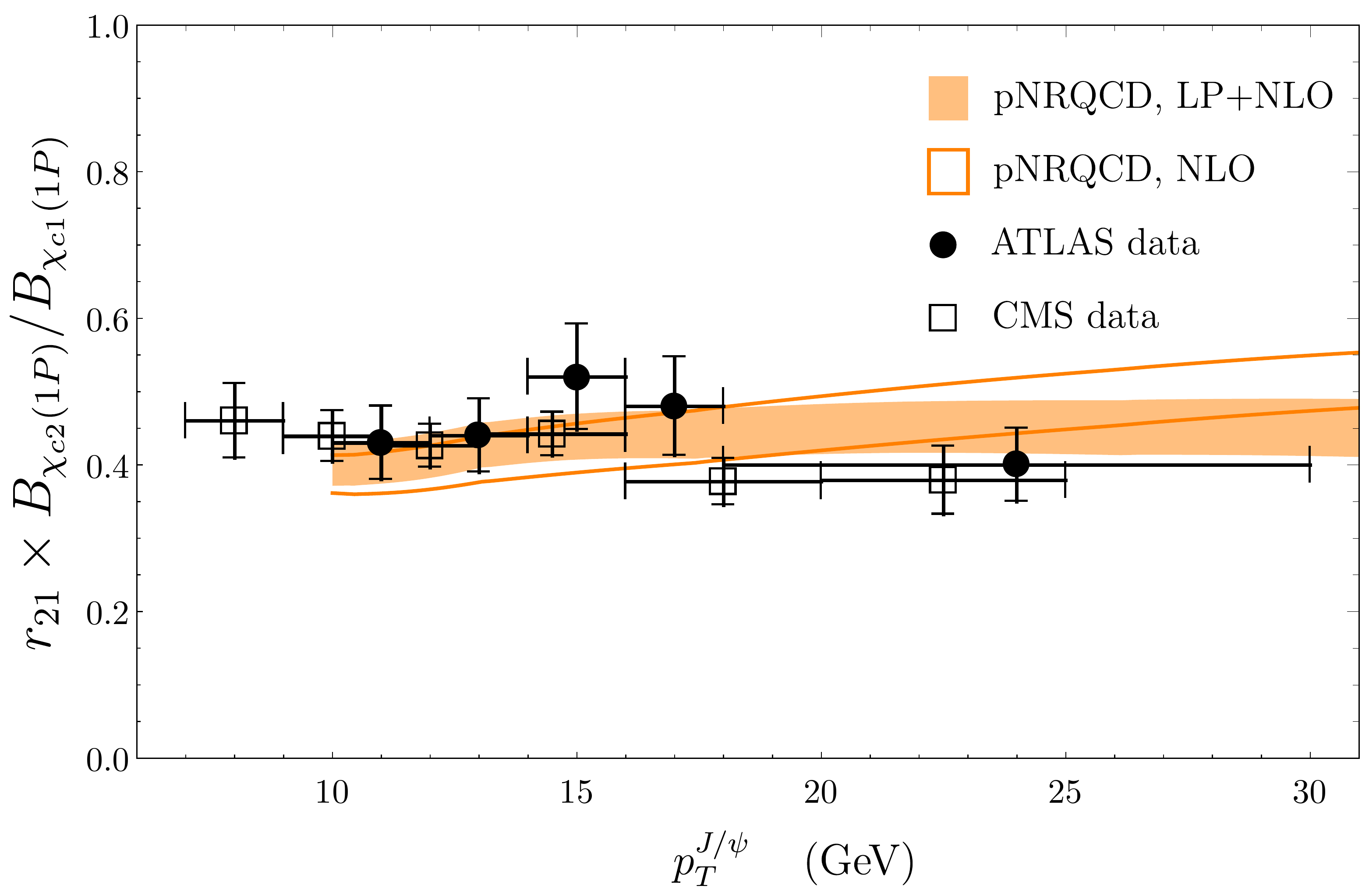}
\caption{
pNRQCD results for the 
cross section ratio $r_{21} \times B_{\chi_{c2}(1P)}/B_{\chi_{c1}(1P)}$ at the
$\sqrt{s}=7$~TeV LHC in the rapidity region $|y|<0.75$ compared to
CMS~\cite{Chatrchyan:2012ub} and
ATLAS~\cite{ATLAS:2014ala} measurements. From ref.~\cite{Brambilla:2021abf}. }
\label{fig:chicratio}
\end{figure}

Since a lattice QCD calculation of ${\cal E}$ has not yet been done, we 
determine ${\cal E}$ by comparing the cross sections computed from 
eq.~(\ref{eq:pwave_fac}) with the measured 
cross section ratio $r_{21} = 
(d \sigma_{\chi_{c2} (1P)}/d p_T)/(d \sigma_{\chi_{c1} (1P)}/d p_T)$ at the
LHC, where $p_T$ is the transverse momentum of the $\chi_{cJ}$.
Note that the ratio is independent of the radial wavefunction. 
We obtain~\cite{Brambilla:2021abf}
\begin{equation}
\label{eq:ENLO}
{\cal E}(\Lambda = 1.5~{\rm GeV})|_{\rm NLO} = 1.17 \pm 0.05.
\end{equation}
Since the fixed-order calculations of the SDCSs may
contain large logarithms of $p_T/m$, resummation of the logarithms can have
significant effects on their shapes in $p_T$. If we 
use the SDCSs from ref.~\cite{Bodwin:2014gia,
Bodwin:2015iua} which include resummed leading logarithms of $p_T/m$ at
leading power (LP) in $m/p_T$, we obtain~\cite{Brambilla:2021abf} 
\begin{equation}
\label{eq:ELPNLO}
{\cal E}(\Lambda = 1.5~{\rm GeV})|_{\rm LP+NLO} = 4.48 \pm 0.14.
\end{equation}
We compare the pNRQCD result for 
$r_{21} \times B_{\chi_{c2}(1P)}/B_{\chi_{c1}(1P)}$
obtained in ref.~\cite{Brambilla:2021abf}
as a function of the transverse momentum $p_T^{J/\psi}$ of the $J/\psi$ produced
in decays of $\chi_{cJ}$, 
where $B_{\chi_{cJ}(1P)} =
{\rm Br} [\chi_{cJ}(1P) \to J/\psi \gamma] \times {\rm Br} (J/\psi \to \mu^+
\mu^-)$, with measurements at the LHC in figure~\ref{fig:chicratio}. 

\begin{figure}[h]
\centering
\includegraphics[width=7cm,clip]{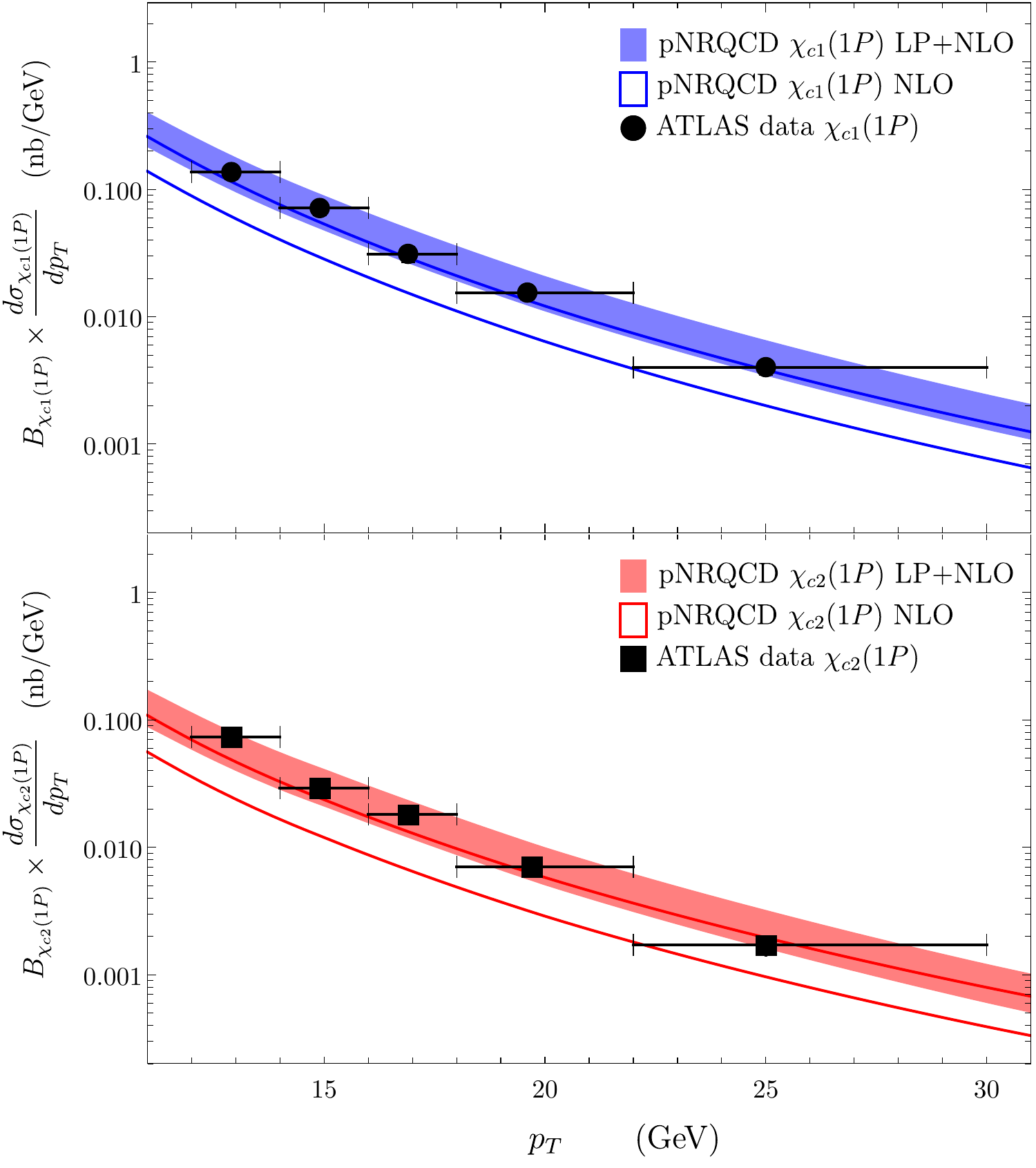}
\caption{
pNRQCD results for the 
production cross sections of $\chi_{c1}(1P)$ and $\chi_{c2} (1P)$ at the 
$\sqrt{s}=7$~TeV LHC in the rapidity region $|y|<0.75$ 
compared to ATLAS~\cite{ATLAS:2014ala} measurements. 
From ref.~\cite{Brambilla:2021abf}. }
\label{fig:chicrate}
\end{figure}

Based on the results for ${\cal E}$ that we obtain, we can compute absolute
cross sections of $\chi_{cJ}$ at the LHC. We use 
$|R^{(0)}{}'(0)|^2 = 0.057$~GeV$^5$, which is determined 
in ref.~\cite{Brambilla:2021abf} 
from the two-photon widths of $\chi_{c1}$ and $\chi_{c2}$ at one-loop level and
measurements in ref.~\cite{Ablikim:2012xi}. 
We show the pNRQCD results for the hadroproduction cross sections 
at the LHC from ref.~\cite{Brambilla:2021abf} compared 
to ATLAS~\cite{ATLAS:2014ala} measurements in figure~\ref{fig:chicrate}. 

\begin{figure}[h]
\centering
\includegraphics[width=8cm,clip]{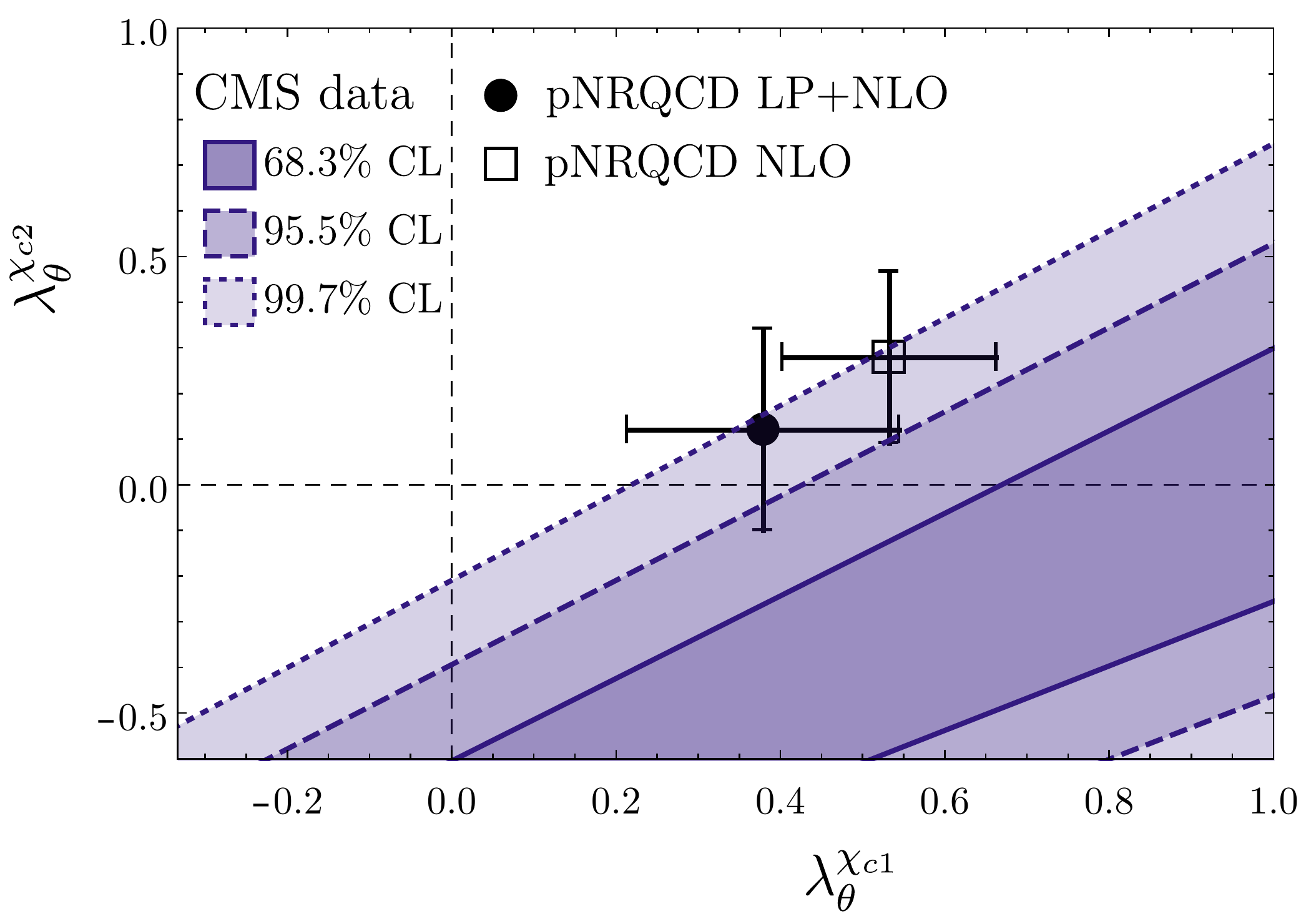}
\caption{
pNRQCD results for the 
polarization parameters $\lambda_\theta^{\chi_{cJ}}$ for $J=1$ and 2 at the
$\sqrt{s} =7$~TeV LHC averaged over 8~GeV$< p_T^{J/\psi} <$ 30~GeV compared
with the experimental constraints from CMS~\cite{Sirunyan:2019apc}. 
From ref.~\cite{Brambilla:2021abf}. }
\label{fig:chicpol}
\end{figure}

Finally, we consider the polarization of the $J/\psi$ from decays of
$\chi_{c1}(1P)$ and $\chi_{c2}(1P)$. The polarization parameter
$\lambda_\theta^{\chi_{cJ}}$ is defined by 
$\lambda_\theta^{\chi_{cJ}} = (1-3 \xi_{\chi_{cJ}})/ (1+\xi_{\chi_{cJ}})$, 
where $\xi_{\chi_{cJ}}$ is the fraction of longitudinally produced $J/\psi$ 
from decays of $\chi_{cJ}(1P)$. We compare in figure~\ref{fig:chicpol} 
the pNRQCD results for 
$\lambda_\theta^{\chi_{c1}}$ and $\lambda_\theta^{\chi_{c2}}$ 
in ref.~\cite{Brambilla:2021abf}
with the experimental constraints from CMS~\cite{Sirunyan:2019apc}. 

\subsection{\boldmath Inclusive hadroproduction of $\chi_{bJ}$}

\begin{figure}[h]
\centering
\includegraphics[width=8cm,clip]{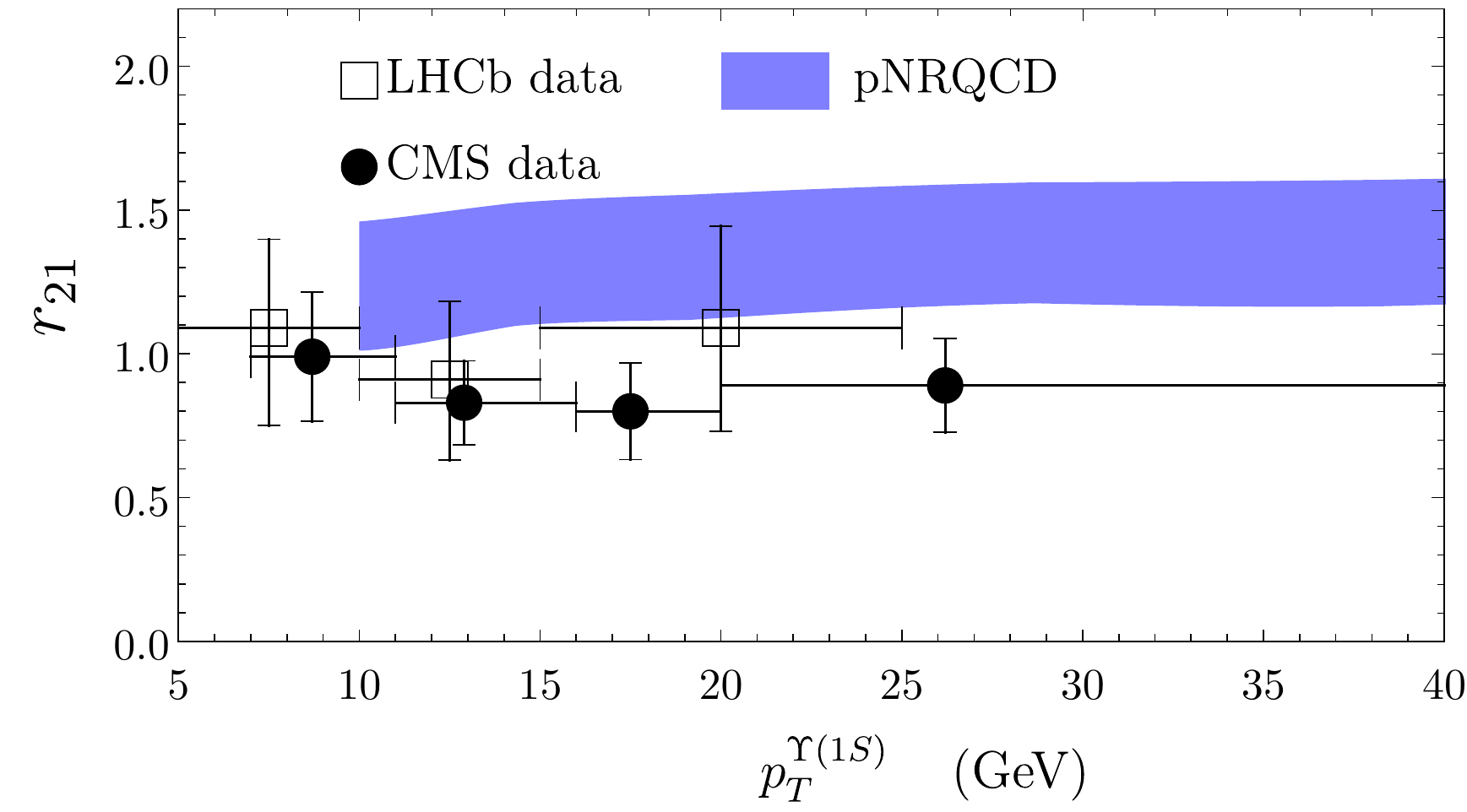}
\caption{
pNRQCD results for the 
cross section ratio $r_{21}$ for the $\chi_{b1}(1P)$ and $\chi_{b2}(1P)$ states
at the $\sqrt{s}=7$~TeV LHC compared with LHCb and CMS data. 
From ref.~\cite{Brambilla:2021abf}. }
\label{fig:chibratio}
\end{figure}

We now compute the hadroproduction rates of $\chi_{bJ}(nP)$ at the LHC, where 
$J=1$, 2 and $n=1$, 2, and 3. 
Similarly to the case of $\chi_{cJ}$, 
we employ the SDCSs computed at 
next-to-leading order
accuracy in $\alpha_s$ from ref.~\cite{Wan:2014vka}, and take 
$\Lambda=4.75$~GeV for the $\overline{\rm MS}$ scale associated with the 
color-octet LDME.
We compute ${\cal E}(\Lambda= 4.75$~GeV$)$ from 
${\cal E}(\Lambda_0=1.5~{\rm GeV})$ by using the one-loop renormalization-group
improved formula 
\begin{equation}
{\cal E}(\Lambda) = 
{\cal E}(\Lambda_0) + \frac{24 C_F}{\beta_0}
\log \frac{\alpha_s(\Lambda_0)}{\alpha_s (\Lambda)}, 
\end{equation}
where $\beta_0 = 11 N_c/3-2 n_f/3$ and $n_f=4$ is the number of active quark
flavors. We take the average of eq.~(\ref{eq:ENLO}) and eq.~(\ref{eq:ELPNLO})
to obtain 
${\cal E}(\Lambda_0=1.5~{\rm GeV}) = 2.8 \pm 1.7$. 
From this we can compute the ratio $r_{21}$ of the $\chi_{b1}(1P)$ and
$\chi_{b2}(1P)$, which is independent of the 
radial wavefunction. We compare the pNRQCD result for the ratio $r_{21}$ 
from ref.~\cite{Brambilla:2021abf} as a function of 
the transverse momentum $p_T^{\Upsilon(1S)}$ of
the $\Upsilon(1S)$ from decays of $\chi_{bJ}$
with LHCb~\cite{Aaij:2014hla} and CMS~\cite{Khachatryan:2014ofa} 
measurements in figure~\ref{fig:chibratio}.

\begin{figure}[h]
\centering
\includegraphics[width=5.2cm,clip]{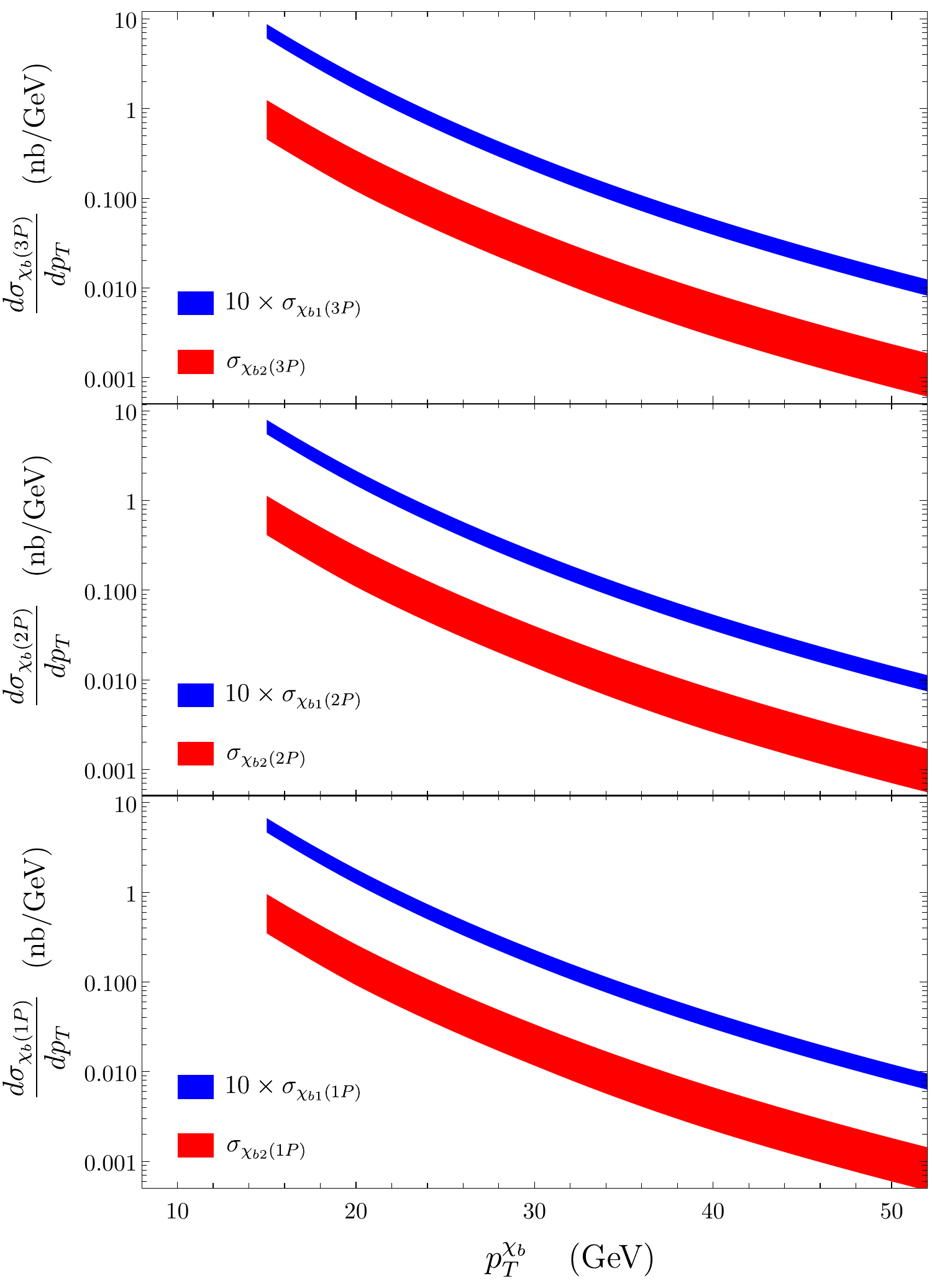}
\includegraphics[width=7.3cm,clip]{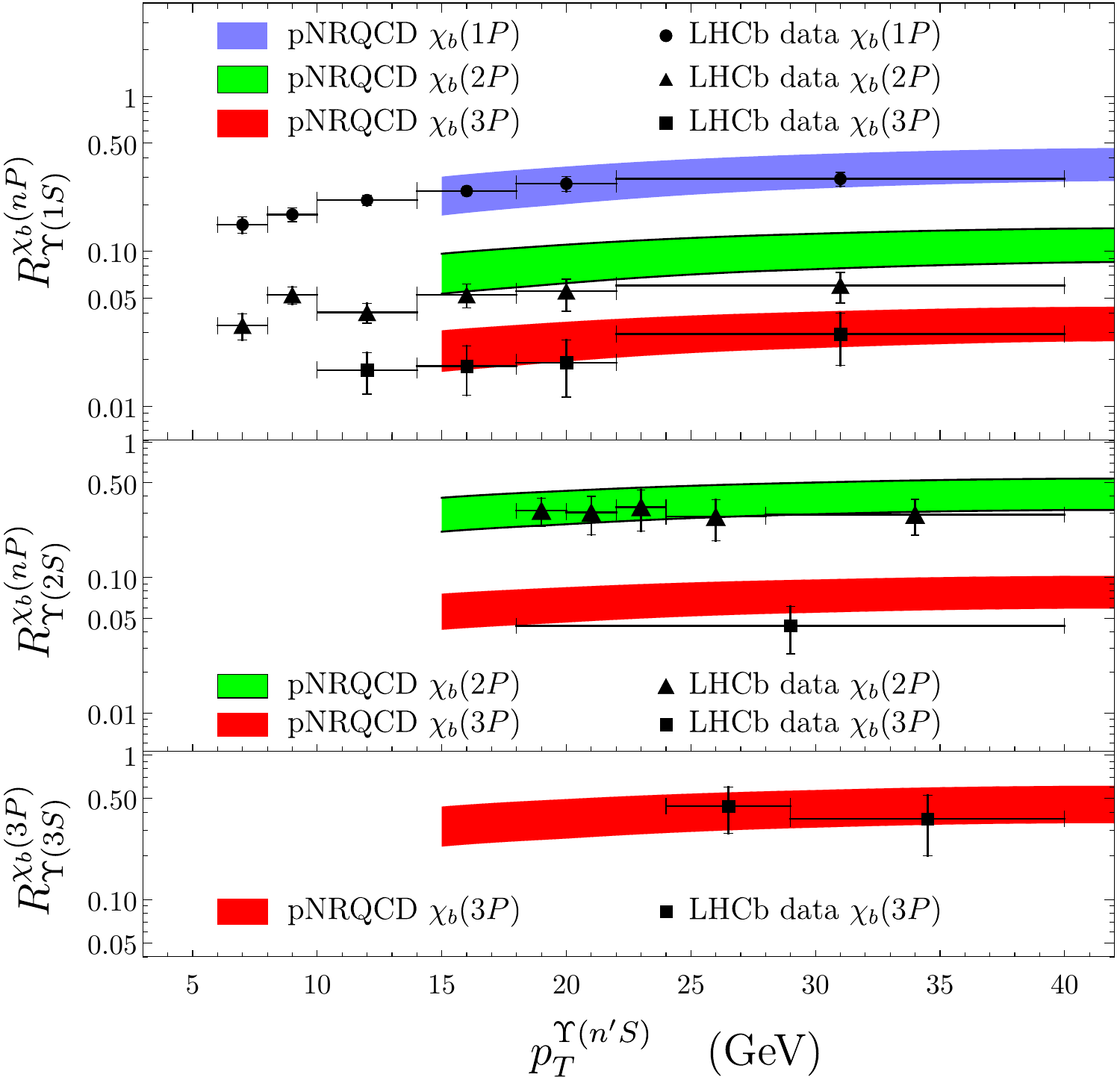}
\caption{
Left panel : pNRQCD results for the inclusive production rates of
$\chi_{bJ}(nP)$ ($J=1$, 2 and $n=1$, 2, and 3) at the $\sqrt{s}=7$~TeV LHC in
the rapidity range $2 < y < 4.5$. 
Right panel : pNRQCD results for the feeddown fractions 
$R_{\Upsilon(n'S)}^{\chi_b(nP)}$ compared to LHCb~\cite{Aaij:2014caa} data. 
From ref.~\cite{Brambilla:2021abf}. }
\label{fig:chibrate}
\end{figure}

In order to compute absolute cross sections of $\chi_{bJ}(nP)$, 
we take the values for $|R^{(0)}{}'(0)|^2$ from the averages of potential-model
calculations considered in ref.~\cite{Brambilla:2020xod}, 
which are given by $|R^{(0)}_{1P}{}'(0)|^2 = 1.47$~GeV$^5$, 
$|R^{(0)}_{2P}{}'(0)|^2 = 1.74$~GeV$^5$, and 
$|R^{(0)}_{3P}{}'(0)|^2 = 1.92$~GeV$^5$. 
The results obtained in ref.~\cite{Brambilla:2021abf} for 
the absolute cross sections of 
$\chi_{b1}(nP)$ and $\chi_{b2}(nP)$ for $n=1$, 2, and 3 are shown in the left
panel of figure~\ref{fig:chibrate}. 
In order to compare with available measurements from LHCb~\cite{Aaij:2014caa}, 
we compute the feeddown
fractions $R_{\Upsilon(n'S)}^{\chi_b(nP)}$ for $n \ge n'$, 
which are the fractions of
$\Upsilon(n'S)$ produced from decays of $\chi_b(nP)$. 
The fractions $R_{\Upsilon(n'S)}^{\chi_b(nP)}$ are computed from the 
$\chi_{cJ} (nP)$ production cross sections multiplied by 
the branching ratios ${\rm Br}[\chi_{cJ}(nP) \to \Upsilon(n'S)+\gamma]$,
divided by the the inclusive production rate of $\Upsilon(n'S)$. 
We take the direct $\Upsilon(n'S)$ production cross sections computed in
ref.~\cite{Han:2014kxa}, 
from which we compute the inclusive production rates by adding the feeddown
contributions from decays of $\chi_b(nP)$ and $\Upsilon(n''S)$, where 
$n \ge n'$ and $n'' \ge n'+1$. 
We show the results for $R_{\Upsilon(n'S)}^{\chi_b(nP)}$ at the $\sqrt{s}
=7$~TeV LHC from ref.~\cite{Brambilla:2021abf}
compared with LHCb~\cite{Aaij:2014caa} data in the right panel of 
figure~\ref{fig:chibrate}. 

\section{Summary and outlook}

We reviewed recent results for inclusive hadroproduction of $P$-wave heavy
quarkonia in potential NRQCD. By working in the strong coupling regime,  
expressions for the NRQCD LDMEs for inclusive production of heavy quarkonia
have been obtained in terms of quarkonium wavefunctions and universal gluonic
correlators. This greatly reduces the number of nonperturbative unknowns,
thanks to the universal nature of the gluonic correlators. 
In the case of $P$-wave heavy quarkonia, the LDMEs that appear in NRQCD
factorization formulas at leading order in $v$ can be determined from the
quarkonium wavefunctions and a single gluonic correlator ${\cal E} = 
\frac{3}{N_c} \delta^{ij} {\cal E}^{ij}$, where the tensor ${\cal E}^{ij}$ is
defined in eq.~(\ref{eq:Etensor}). 
Based on this result, the hadroproduction cross sections of $\chi_{cJ}$ and 
$\chi_{bJ}$ have been computed in refs.~\cite{Brambilla:2020ojz,
Brambilla:2021abf} by using a determination of ${\cal E}$ 
from measurements of the ratio of $\chi_{c1}$ and $\chi_{c2}$ cross sections at 
the LHC. The phenomenological results are in reasonable agreement with
available LHC data. 

The formalism developed in ref.~\cite{Brambilla:2021abf} for the 
LDMEs can be applied to inclusive 
production rates of any strongly coupled heavy quarkonium state. 
It would especially be interesting 
to compute in potential NRQCD the production LDMEs for $S$-wave quarkonia 
such as $J/\psi$, $\psi(2S)$, and $\Upsilon$, which may shed light on the
long-standing puzzle of $J/\psi$ polarization and help understand the 
$\eta_c$ production mechanism~\cite{LHCb:2014oii, Butenschoen:2014dra, 
Han:2014jya, Zhang:2014ybe, LHCb:2019zaj}.

\bibliography{hadropnrqcd_proceeding.bib}

\begin{thebibliography}{36}

\bibitem{Brambilla:2004wf}
N.~Brambilla et~al. (Quarkonium Working Group) (2004), \texttt{hep-ph/0412158}

\bibitem{Brambilla:2010cs}
N.~Brambilla et~al., Eur. Phys. J. C \textbf{71}, 1534 (2011),
  \texttt{1010.5827}

\bibitem{Bodwin:2013nua}
G.T. Bodwin, E.~Braaten, E.~Eichten, S.L. Olsen, T.K. Pedlar, J.~Russ,
  \emph{{Quarkonium at the Frontiers of High Energy Physics: A Snowmass White
  Paper}}, in \emph{{Community Summer Study 2013}: {Snowmass on the
  Mississippi}} (2013), \texttt{1307.7425}

\bibitem{Brambilla:2014jmp}
N.~Brambilla et~al., Eur. Phys. J. C \textbf{74}, 2981 (2014),
  \texttt{1404.3723}

\bibitem{Caswell:1985ui}
W.E. Caswell, G.P. Lepage, Phys. Lett. \textbf{167B}, 437 (1986)

\bibitem{Bodwin:1994jh}
G.T. Bodwin, E.~Braaten, G.P. Lepage, Phys. Rev. \textbf{D51}, 1125 (1995),
  [Erratum: Phys. Rev.D55,5853(1997)], \texttt{hep-ph/9407339}

\bibitem{Chung:2018lyq}
H.S. Chung, PoS \textbf{Confinement2018}, 007 (2018), \texttt{1811.12098}

\bibitem{Brambilla:1999xf}
N.~Brambilla, A.~Pineda, J.~Soto, A.~Vairo, Nucl. Phys. \textbf{B566}, 275
  (2000), \texttt{hep-ph/9907240}

\bibitem{Pineda:1997bj}
A.~Pineda, J.~Soto, Nucl. Phys. B Proc. Suppl. \textbf{64}, 428 (1998),
  \texttt{hep-ph/9707481}

\bibitem{Brambilla:2004jw}
N.~Brambilla, A.~Pineda, J.~Soto, A.~Vairo, Rev. Mod. Phys. \textbf{77}, 1423
  (2005), \texttt{hep-ph/0410047}

\bibitem{Brambilla:2001xy}
N.~Brambilla, D.~Eiras, A.~Pineda, J.~Soto, A.~Vairo, Phys. Rev. Lett.
  \textbf{88}, 012003 (2002), \texttt{hep-ph/0109130}

\bibitem{Brambilla:2002nu}
N.~Brambilla, D.~Eiras, A.~Pineda, J.~Soto, A.~Vairo, Phys. Rev. D \textbf{67},
  034018 (2003), \texttt{hep-ph/0208019}

\bibitem{Brambilla:2020xod}
N.~Brambilla, H.S. Chung, D.~M{\"u}ller, A.~Vairo, JHEP \textbf{04}, 095
  (2020), \texttt{2002.07462}

\bibitem{Chung:2020zqc}
H.S. Chung, JHEP \textbf{12}, 065 (2020), \texttt{2007.01737}

\bibitem{Chung:2021efj}
H.S. Chung, JHEP \textbf{09}, 195 (2021), \texttt{2106.15514}

\bibitem{Brambilla:2020ojz}
N.~Brambilla, H.S. Chung, A.~Vairo, Phys. Rev. Lett. \textbf{126}, 082003
  (2021), \texttt{2007.07613}

\bibitem{Brambilla:2021abf}
N.~Brambilla, H.S. Chung, A.~Vairo, JHEP \textbf{09}, 032 (2021),
  \texttt{2106.09417}

\bibitem{Nayak:2005rt}
G.C. Nayak, J.W. Qiu, G.F. Sterman, Phys. Rev. D \textbf{72}, 114012 (2005),
  \texttt{hep-ph/0509021}

\bibitem{Nayak:2005rw}
G.C. Nayak, J.W. Qiu, G.F. Sterman, Phys. Lett. B \textbf{613}, 45 (2005),
  \texttt{hep-ph/0501235}

\bibitem{Nayak:2006fm}
G.C. Nayak, J.W. Qiu, G.F. Sterman, Phys. Rev. D \textbf{74}, 074007 (2006),
  \texttt{hep-ph/0608066}

\bibitem{Wan:2014vka}
L.P. Wan, J.X. Wang, Comput. Phys. Commun. \textbf{185}, 2939 (2014),
  \texttt{1405.2143}

\bibitem{Bodwin:2014gia}
G.T. Bodwin, H.S. Chung, U.R. Kim, J.~Lee, Phys. Rev. Lett. \textbf{113},
  022001 (2014), \texttt{1403.3612}

\bibitem{Bodwin:2015iua}
G.T. Bodwin, K.T. Chao, H.S. Chung, U.R. Kim, J.~Lee, Y.Q. Ma, Phys. Rev. D
  \textbf{93}, 034041 (2016), \texttt{1509.07904}

\bibitem{Chatrchyan:2012ub}
S.~Chatrchyan et~al. (CMS), Eur. Phys. J. C \textbf{72}, 2251 (2012),
  \texttt{1210.0875}

\bibitem{ATLAS:2014ala}
G.~Aad et~al. (ATLAS), JHEP \textbf{07}, 154 (2014), \texttt{1404.7035}

\bibitem{Ablikim:2012xi}
M.~Ablikim et~al. (BESIII), Phys. Rev. D \textbf{85}, 112008 (2012),
  \texttt{1205.4284}

\bibitem{Sirunyan:2019apc}
A.M. Sirunyan et~al. (CMS), Phys. Rev. Lett. \textbf{124}, 162002 (2020),
  \texttt{1912.07706}

\bibitem{Aaij:2014hla}
R.~Aaij et~al. (LHCb), JHEP \textbf{10}, 088 (2014), \texttt{1409.1408}

\bibitem{Khachatryan:2014ofa}
V.~Khachatryan et~al. (CMS), Phys. Lett. B \textbf{743}, 383 (2015),
  \texttt{1409.5761}

\bibitem{Aaij:2014caa}
R.~Aaij et~al. (LHCb), Eur. Phys. J. C \textbf{74}, 3092 (2014),
  \texttt{1407.7734}

\bibitem{Han:2014kxa}
H.~Han, Y.Q. Ma, C.~Meng, H.S. Shao, Y.J. Zhang, K.T. Chao, Phys. Rev. D
  \textbf{94}, 014028 (2016), \texttt{1410.8537}

\bibitem{LHCb:2014oii}
R.~Aaij et~al. (LHCb), Eur. Phys. J. C \textbf{75}, 311 (2015),
  \texttt{1409.3612}

\bibitem{Butenschoen:2014dra}
M.~Butenschoen, Z.G. He, B.A. Kniehl, Phys. Rev. Lett. \textbf{114}, 092004
  (2015), \texttt{1411.5287}

\bibitem{Han:2014jya}
H.~Han, Y.Q. Ma, C.~Meng, H.S. Shao, K.T. Chao, Phys. Rev. Lett. \textbf{114},
  092005 (2015), \texttt{1411.7350}

\bibitem{Zhang:2014ybe}
H.F. Zhang, Z.~Sun, W.L. Sang, R.~Li, Phys. Rev. Lett. \textbf{114}, 092006
  (2015), \texttt{1412.0508}

\bibitem{LHCb:2019zaj}
R.~Aaij et~al. (LHCb), Eur. Phys. J. C \textbf{80}, 191 (2020),
  \texttt{1911.03326}

\end{thebibliography}

\end{document}